\documentclass[preprint,showpacs,preprintnumbers,amsmath,amssymb]{revtex4}
\usepackage{graphicx}% Include figure files
\usepackage{dcolumn}% Align table columns on decimal point
\usepackage{bm}% bold math

\begin{document}

\title{Natural orbits of atomic Cooper pairs in a nonuniform Fermi gas}

% Force line breaks with \\

\author{Y. H. Pong and C. K. Law}
\affiliation{Department of Physics and Institute of Theoretical
Physics, The Chinese university of Hong Kong, Shatin, Hong Kong SAR,
China}

\date{\today}
\begin{abstract}
We examine the basic mode structure of atomic Cooper pairs in an
inhomogeneous Fermi gas. Based on the properties of Bogoliubov
quasi-particle vacuum, the single particle density matrix and the
anomalous density matrix share the same set of eigenfunctions. These
eigenfunctions correspond to natural pairing orbits associated with
the BCS ground state. We investigate these orbits for a Fermi gas in
a spherical harmonic trap, and construct the wave function of a
Cooper pair in the form of Schmidt decomposition. The issue of
spatial quantum entanglement between constituent atoms in a pair is
addressed.

\end{abstract}


\maketitle

\section{Introduction}

Quantum degenerate gas of two-component Fermi atoms provides a
well-controlled system for studying Cooper pairing responsible for
superfluidity phenomena in the BCS regime \cite{stoof,jin}. With the
advances in cooling and trapping techniques, recent experiments have
demonstrated various consequences of pairing
\cite{chin,vortex,rf,shot}. One of the major tasks for theoretical
investigation is the structure of Cooper pairs in non-uniform finite
systems. Since the weakly interacting ultra-cold atoms mainly
involve two-body \textit{s}-wave scattering processes, detailed
analysis can be performed via the pseudo-potential method and
Bogoliubov-deGennes (BdG) equations \cite{castin,deGennes}. An
important quantity that can be obtained self-consistently by this
approach is the gap function. Such a quantity measures the pairing
field at a given point in space, but it lacks the information about
correlation between atoms at different spatial points. In order to
gain a more complete picture of pairing, it is useful to examine
properties of two-point correlation functions \cite{corr}.

In this paper we examine the underlying mode structures inherent in
two-point correlation functions: $\langle
\psi_\alpha^\dag(\textbf{r}_1) \psi_\alpha(\textbf{r}_2)\rangle$ and
$\left\langle \psi_\beta(\textbf{r}_1) \psi_\alpha(\textbf{r}_2)
\right\rangle$, with $\psi_{j} $ ($j=\alpha, \beta$) being the field
operator associated with the spin component $j$. For symmetric
systems, we employ the fact that these correlation functions are
built up by the {\em same} set of orthogonal eigenfunctions
\cite{dobaczewski}. Such eigenfunctions can be interpreted as
natural pairing orbits that form the BCS wave functions in
inhomogeneous systems. We note that the standard textbook
description of BCS wave function refers to infinite homogeneous
systems such that each pairing orbit corresponds to the
eigenfunctions of opposite momenta \cite{deGennes}. However, the
presence of a confining potential in inhomogeneous systems could
drastically alter the pairing orbits. Although the time-reversal
symmetry helps to fix a certain set of quantum numbers in pairs, the
exact form of pairing orbits are difficult to find in general. One
possible solution is to treat the pairing orbits as (unknown)
variational functions. With the usual BCS ansatz and variation
technique, a set of nonlinearly coupled equations of pairing modes
have been derived \cite{dobaczewski,German}. However, these
nonlinear equations have no simple physical interpretations, and
numerical solutions have only been demonstrated in nuclear systems
\cite{dobaczewski,German}. As the particle numbers in Fermi gases
systems are typically much larger than that in nuclei, numerical
method based on variational method could become difficult. To our
knowledge, natural orbits of Cooper pairs in a trapped Fermi atomic
gases have not been fully explored.

One of the main purposes of our paper is to indicate an efficient
way of determining pairing orbits, directly from the two-point
correlation functions obtained by BdG mean field equations.  To
illustrate our method, we will examine the gas in a spherical
harmonic trap. For such systems, Bruun and Heiselberg have provided
useful insight about the approximate forms of pairing orbits with a
different approach \cite{bruun,heiselberg}. Here, we will present
numerically exact examples of pair orbits. With these pairing orbits
we can further construct and study the spatial wave function of a
Cooper pair. In addition, we will address the quantum entanglement
between two constituent atoms in a pair. Recently, one of us have
addressed the issue of quantum entanglement of two atoms due to
$s$-wave scattering \cite{WJ}. The study of a Cooper pair would shed
some light on the importance of many-body effects.

\section{Theory}

\subsection{The model and BdG equations}
To begin with, we write down the model Hamiltonian describing an
interacting two-component Fermi gas in a trapping potential
$U_0(\textbf{r})$,
\begin{eqnarray} \label{eq:HAMILTONIAN_PRIMITIVE}
H =\int d^3 {\bf r} \; \left[{\psi_{\alpha}^{\dag}(\textbf{r}) H_0
\psi_{\alpha}(\textbf{r}) + \psi_{\beta}^{\dag}(\textbf{r}) H_0
\psi_{\beta}(\textbf{r})} \right]  \nonumber \\
+ \;g \int d^3 {\bf r} \;
\psi_{\alpha}^{\dag}(\textbf{r})\psi_{\beta}^{\dag}(\textbf{r})
\psi_{\beta}(\textbf{r})\psi_{\alpha}(\textbf{r})
\end{eqnarray}
where $H_0=-\frac{\hbar^{2}}{2m}\nabla^{2}+U_0(\textbf{r})-\mu$,
with $m$ and $\mu$ being the particle mass and chemical potential
respectively. The coupling strength $g$ is related to the $s$-wave
scattering length $a$ via $g=4\pi \hbar^2 a / m$. In this paper we
assume that $a$ is negative and the number of atoms in each
component is the same.

Under the mean field approximation, the Hamiltonian can be
diagonalized through the Bogoliubov transformation:
$\psi_{\alpha}(\textbf{r})=\sum\nolimits_{\eta}u_{\eta}(\textbf{r})
\gamma_{\eta\alpha}-v_{\eta}^{*}(\textbf{r}) \gamma_{\eta
\beta}^\dagger$, and $
\psi_{\beta}(\textbf{r})=\sum\nolimits_{\eta}u_{\eta}
(\textbf{r})\gamma_{\eta \beta}+v_{\eta}^{*}(\textbf{r})
\gamma_{\eta \alpha}^\dagger $, such that the ground state is the
vacuum state of the quasi-particle operators $\gamma_{\eta}$'s,
i.e., Bogoliubov vacuum. The quasi-particle wave functions
$u_{\eta}(\textbf{r})$ and $v_{\eta}(\textbf{r})$ are solved
self-consistently by the Bogoliubov-deGennes (BdG) equations
\cite{deGennes}:
\begin{eqnarray}
\begin{array} {ccc}
\left[ H_0(\textbf{r})+W(\textbf{r})\right] u_{\eta}(\textbf{r})
+ \Delta(\textbf{r})v_{\eta}(\textbf{r})&=&E_{\eta}
u_{\eta}(\textbf{r}) \\
\Delta^{*}(\textbf{r})u_{\eta}(\textbf{r})
-\left[H_0(\textbf{r})+W(\textbf{r})\right]v_{\eta}(\textbf{r})
&=&E_{\eta}v_{\eta}(\textbf{r}) \label{eq:BDG_1}
\end{array}
\end{eqnarray}
with $W(\textbf{r})=g \sum\nolimits_{\eta}|v_{\eta}(\textbf{r})|^2$,
and $\Delta(\textbf{r})=-g
\sum\nolimits_{\eta}u_{\eta}(\textbf{r})v_{\eta}^*(\textbf{r})$. It
is important to note that $\Delta(\textbf{r})$ has a $1/r$
divergence \cite{castin,Yu}. Such a divergence can be eliminated by
replacing $g$ with a regularized effective coupling constant as
described in Ref. \cite{Yu,Urban}.

\subsection{Two-point correlation functions}
Given that the ground state of the system is described by the
Bogoliubov quasi-particle vacuum, the normal density matrix
$\rho(\textbf{r}_1,\textbf{r}_2)=\left\langle
\psi_\alpha^\dag(\textbf{r}_1)\psi_\alpha(\textbf{r}_2)\right
\rangle=\langle
\psi_\beta^\dag(\textbf{r}_1)\psi_\beta(\textbf{r}_2)\rangle$ and
the anomalous density matrix $\nu(\textbf{r}_1,\textbf{r}_2)=\langle
\psi_\beta(\textbf{r}_1)\psi_\alpha(\textbf{r}_2)\rangle$ take the
form:
\begin{eqnarray}
&& \rho(\textbf{r}_1,\textbf{r}_2)=\sum\nolimits_\eta v_{\eta}
    (\textbf{r}_1)v_{\eta}^*(\textbf{r}_2) \label{eq:RHO_BDG}\\
&& \nu(\textbf{r}_1,\textbf{r}_2)=-\sum\nolimits_\eta u_{\eta}
    (\textbf{r}_1)v_{\eta}^*(\textbf{r}_2) .\label{eq:NU_BDG}
\end{eqnarray}
In this paper we assume that $\rho(\textbf{r}_1,\textbf{r}_2)$ and
$\nu(\textbf{r}_1,\textbf{r}_2)$ are real symmetric matrices. Such a
symmetric property can be shown explicitly in spherically trapped
systems that we will discussed later in the paper.

From the properties of quasi-particle wave functions, it can be
shown that $\rho$ and $\nu$ commute (see Appendix), i.e.,
\begin{equation} \label{eq:COMMUTE}
\int d^3r_1 \; \left[{\rho(\textbf{r},\textbf{r}_1)
\nu(\textbf{r}_1,\textbf{r}_2) -\nu(\textbf{r},\textbf{r}_1)
\rho(\textbf{r}_1,\textbf{r}_2)} \right] =0.
\end{equation}
This implies that the normal density matrix and the anomalous density
matrix share a common set of eigenfunctions defined by the integral
equations:
\begin{eqnarray}
 \int {d^3 r_2 \; \rho(\textbf{r}_1,\textbf{r}_2)} \; f_n (\textbf{r}_2)
 = \lambda_n f_n (\textbf{r}_1) \label{eq:EIGEN_W}\\
 \int {d^3 r_2 \; \nu(\textbf{r}_1,\textbf{r}_2)}  \; f_n (\textbf{r}_2)
 = \chi_n f_n (\textbf{r}_1) \label{eq:EIGEN_PAIR}
\end{eqnarray}
where $\lambda_n$ and $\chi_n$ are real eigenvalues, and $ \left\{
{f_n } \right\}$ is a complete and orthogonal set of eigenfunctions.
Furthermore, $\rho(\textbf{r}_1,\textbf{r}_2)$ and
$\nu(\textbf{r}_1,\textbf{r}_2)$ can be expressed as the following
bilinear expansion:
\begin{eqnarray}
 \left\langle {\psi_\alpha^\dag (\textbf{r}_1)\psi_\alpha (\textbf{r}_2)}
 \right\rangle = \sum\nolimits_n { \lambda_n f_n(\textbf{r}_1)
 f_n^* (\textbf{r}_2)}
\label{eq:W_BCS1}\\
 \left\langle {\psi_\beta (\textbf{r}_1 )\psi_\alpha (\textbf{r}_2)}
 \right\rangle = \sum\nolimits_n {\chi_n f_{n} (\textbf{r}_1) f_n^*
 (\textbf{r}_2)}.
\label{eq:PAIRING_BCS1}
\end{eqnarray}
Here the convergence is ensured by the squared integrable property
of the kernels, according to the theory of integral equations.

\subsection{Natural orbits}

The eigenfunctions $f_n$ are called natural orbits. To interpret
such orbits, let us write down a general mode expansion of operators
$ \psi_\alpha$ and $\psi_\beta$ such that: $ \psi_\alpha
(\textbf{r})=\sum\nolimits_n {\phi_n^{(\alpha)} (\textbf{r})\alpha_n
}$, $\psi_\beta (\textbf{r})=\sum\nolimits_n {\phi_n^{(\beta)}
(\textbf{r})\beta_n }$, where $ \{ \phi_n^{(\alpha)} \}$ and $ \{ {
\phi_n^{(\beta)}} \}$ are two sets of complete orthogonal functions
to be determined, and $\alpha_n$ and $\beta_n$ are the corresponding
fermion annihilation operators. Next we construct the BCS ground
state in the standard form:
\begin{eqnarray} \label{eq:BCS_ANSATZ}
\left| {\Phi } \right\rangle  = \prod_n {(\tilde u_n + \tilde v_n \alpha
_n^\dag  \beta_{n}^\dag )\left| 0 \right\rangle }
\end{eqnarray}
with the normalized coefficients $\tilde u_n$ and $\tilde v_n$
satisfying $\left|\tilde u_n\right|^2+\left|\tilde v_n\right|^2=1$.
From the fact that
\begin{eqnarray}
\langle \Phi |{\psi_\alpha^\dag (\textbf{r}_1)\psi_\alpha
 (\textbf{r}_2)} |\Phi
\rangle = \sum\nolimits_n
{\phi_n^{(\alpha)*}(\textbf{r}_1)\phi_n^{(\alpha)} (\textbf{r}_2)}
 \left|\tilde v_n\right|^2 \\
\langle \Phi |{\psi_\beta (\textbf{r}_1 )\psi_\alpha (\textbf{r}_2)}
|\Phi  \rangle = \sum\nolimits_n {\phi_{n}^{(\beta)} (\textbf{r}_1)
\phi_n^{(\alpha)} (\textbf{r}_2)} \tilde u_n^* \tilde v_n
\end{eqnarray}
we immediately see that the constructed BCS ground state
(\ref{eq:BCS_ANSATZ}) is consistent with the BdG results Eq.
(\ref{eq:W_BCS1}) and (\ref{eq:PAIRING_BCS1}) if
\begin{eqnarray}
&& \phi_n^{(\alpha)}( {\bf r})= f_n^{*}( {\bf r})\\
&& \phi_n^{(\beta)}( {\bf r})= f_n ({\bf r})
\end{eqnarray}
and $\tilde v_n = \sqrt \lambda_n$ and $\tilde u_n = \chi_n / \sqrt
\lambda_n$ (Appendix). In other words, $f_n$ and its conjugate are
indeed the pairing modes needed for the construction of the BCS
ground state. This is shown by matching the correlation functions
obtained from BdG mean-field equations. In fact, with the help of
Wick's theorem, the choice of mode functions (13) and (14) also
match higher order correlation functions. Hence the equivalence
between Bogoliubov vacuum and the BCS ground state in a trapped
system can be established explicitly.

Historically, the concept of natural orbits were applied to pairing
problems of nucleons and one can find some early references in Ref.
\cite{dobaczewski}. Here we employ such an idea to analyze the
structure of Cooper pairs in atomic systems. We should emphasize
that the key to the existence of $f_n$ is the symmetric property of
$\rho$, which makes $\rho$ and $\nu$ commute. In the appendix we
indicate this point in a simple derivation of Eq. (\ref{eq:COMMUTE}). For
asymmetric systems, such as the systems with imbalance populations
of the two components, $\rho$ and $\nu$ do not commute in general.
This would then forbid the ground state to be in the BCS form,
although the BdG equations could still be used to describe
asymmetric systems.

\subsection{Cooper pair wave function and quantum entanglement}

The description of a Cooper pair is inherited from the mean field
description, where every pair in the system is assumed to be
identical. Let us now construct the wave function of a Cooper pair
based on the pairing orbits. This is achieved by noting that the BCS
state takes the form:
\begin{equation}
|\Phi \rangle \propto \exp ( {\sum_j \kappa_j \alpha_j^{\dag}
\beta_j^{\dag}} )|0 \rangle = \sum_{k=0}^\infty \frac {{A^{\dag}}^
k} {k!} |0 \rangle
\end{equation}
where $\kappa_j=\tilde v_{j} / \tilde u_{j}$, $A^\dag=\sum_j
\kappa_j \alpha_j^{\dag} \beta_j^{\dag}$, and $j$ stands for quantum
numbers collectively. The wave function corresponding to the pair
creation operator $A^\dag$ is given by
\begin{equation}
F(\textbf{r}_1,\textbf{r}_2)=C \sum_{j} \kappa_j
f_{j}(\textbf{r}_1)f_{j}^* (\textbf{r}_2)
\end{equation}
where $C$ is a normalization constant. Such a function is
interpreted as the wave function of a Cooper pair. One reasoning is
based on the fact that the BCS state is dominated by particle
numbers near the mean value $\bar N$ when the particle number is
large, i.e., $k \approx \bar N$ terms in Eq. (15) contribute most.
This allows us to have an approximate picture of $\bar N$ pairs
described by $A^{\dag {\bar N}}|0 \rangle$. It is worth noting that
each term in expansion (16) is weighted by the factor $\kappa_j$,
which is different from Eq. (8) and (9). Since $\kappa_j$ is larger
for the more filled orbits, orbits below the Fermi level could have
more contributions individually.

Eq. (16) is precisely a form of Schmidt decomposition of a bipartite
system \cite{knight}. This is observed by the fact that $f_j$ are
orthogonal basis functions, and $\kappa_j$ are corresponding Schmidt
eigenvalues. Schmidt decomposition provides a useful way to
characterize quantum entanglement between two subsystems in pure
states. In particular the degree of quantum entanglement can be
learned by the von Neumann entropy according to the distribution of
Schmidt eigenvalues. Here we apply this method to address the
entanglement between two atoms in a Cooper pair. A transparent and
direct measure of entanglement is the `average' number of Schmidt
modes involved.  The effective Schmidt number $K$ provides this
average \cite{Grobe,WJ}:
\begin{equation}
K=1/\sum_j |\kappa_j'|^4,
\end{equation}
where $\kappa_j'=C \kappa_j$. Such a quantity is also the inverse of
the purity function. The larger the value of $K$, the higher the
entanglement. Note that the entanglement here refers to the spatial
degrees of freedom, which should be distinguished from the spin
entanglement \cite{dunning}.

\section{Fermi gases in a spherical harmonic potential}

Having described the formalism, we now proceed to study natural
orbits of a Fermi gas confined in a spherical harmonic potential:
$U_{0}(r)=\frac{1}{2}m\omega^{2}r^{2}$, with $\omega$ being the
trapping frequency. First, we determine the quasi-particle mode
functions $u_\eta$'s and $v_\eta$'s numerically from the BdG
equations. The spherical symmetry of the system gives: $ru_{\eta
lm}(\textbf{r})=u_{\eta l}(r)Y_{lm}(\theta,\phi)$, $r v_{\eta
lm}(\textbf{r})=v_{\eta l}(r)Y_{lm}(\theta,\phi)$, with $Y_{lm}$
being the spherical harmonic functions. It should be noted that in
order to remove the ultraviolet divergence, we employ an effective
coupling constant $g_{\textrm{eff}}(\textbf{r})$, such that the
whole set of equations is independent on cutoff. The formulation of
renormalization scheme has been discussed extensively in
\cite{Urban,Yu}, and we will skip the details in this paper.

Once numerical solutions of $u_\eta$'s and $v_\eta$'s are found, the
correlation functions $\rho(\textbf{r}_1,\textbf{r}_2)$ and
$\nu(\textbf{r}_1,\textbf{r}_2)$ are obtained from Eqs.
(\ref{eq:RHO_BDG}) and (\ref{eq:NU_BDG}). A significant
simplification can be made by using the addition formula: $(2l +
1)P_l \left( {\cos \gamma } \right) = 4\pi \sum\limits_{m = - l}^l
{Y_{lm}^{*} \left( {\theta _1 ,\phi _1 } \right)Y_{lm} \left(
{\theta _2 ,\phi _2 } \right)}$, where $P_{l}(\cos\gamma)$ is the
Legendre polynomial, and $\gamma$ is the angle between between
$\textbf{r}_1$ and $\textbf{r}_2$. This gives
\begin{eqnarray}
&& \left\langle {\psi _\alpha ^\dag(\textbf{r}_1)\psi
    _\alpha(\textbf{r}_2 )} \right\rangle
    =\sum\nolimits_l P_{l}(\cos\gamma)\alpha_{l}(r_1,r_2)
    \\
&& \left\langle {\psi_\beta (\textbf{r}_1)\psi
    _\alpha(\textbf{r}_2 )} \right\rangle
    =-\sum\nolimits_l P_{l}(\cos\gamma)\beta_{l}(r_1,r_2)
\end{eqnarray}
where
\begin{eqnarray}
\alpha_{l}(r_1,r_2)=\frac{2l+1}{4\pi}\sum_{\eta}
\frac{v_{\eta l}(r_1)v^*_{\eta l}(r_2)}{r_1 r_2}\\
\beta_{l}(r_1,r_2)=\frac{2l+1}{4\pi}\sum_{\eta}\frac{u_{\eta l}(r_1)
v^*_{\eta l}(r_2)}{r_1 r_2}.
\end{eqnarray}
Therefore the correlation functions (\ref{eq:RHO_BDG}) and (\ref{eq:NU_BDG}) depends only on
three variables: $r_1$, $r_2$ and $\gamma$. The integral eigenvalue
equations (\ref{eq:EIGEN_W}) and (\ref{eq:EIGEN_PAIR}), can be
solved conveniently by expressing the eigenfunctions as
$f_{nlm}(\textbf{r})=\frac{1}{r}h_{nl}(r)Y_{lm}(\theta,\phi)$, with
$h_{nl}(r)$ being the radial function associated with the radial
quantum number $n$ and orbital angular momentum $l$. Equations (6)
and (7) then become,
\begin{eqnarray}
&& \int r_1 r_2 \alpha_l(r_1,r_2)h_{nl}(r_2) dr_2 = |\tilde
v_{nl}|^2 h_{nl}(r_1)
\label{eq:INTEGRAL_W} \\
&& \int r_1 r_2 \beta_l(r_1,r_2)h_{nl}(r_2) dr_2 = \tilde u_{nl}^*
\tilde v_n h_{nl}(r_1) \label{eq:INTEGRAL_PAIR}
\end{eqnarray}
which are one-dimensional integral equations, as only the radial
coordinates are involved.

\begin{figure}
\includegraphics [width=7 cm] {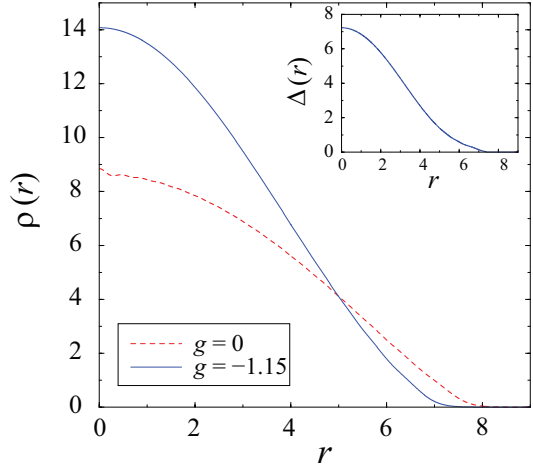}
\caption{\label{fig:DENSITY} (Color online) The \textit{g}= -1.15
(solid line) and the non-interacting (dashed line) radial density
distribution function, normalized to \textit{N}=$\int 4\pi r^2
\rho(r) dr$. Inset shows the pairing field $\Delta(r)$ with
\textit{g}= -1.15.}
\end{figure}

To provide a concrete example, we consider a system of
$N=N_\alpha+N_\beta=2N_\alpha=10912$
particles with the interaction strength $g=-1.15$ in trap units,
(i.e., energy in $\hbar\omega$, length in $\sqrt{\hbar/m\omega}$).
The choice of such a particle number corresponds to the Fermi energy
$\epsilon_F=31.5\hbar\omega$ in the non-interacting limit.
Eq.(\ref{eq:BDG_1}) was solved in a truncated harmonic oscillator
states basis, with a cutoff energy $\sim 180\hbar\omega\gg \mu$. To
achieve convergent results, we employed the regularization scheme
according to Ref. \cite{Urban}. The particle density $\rho(r)$ and
pairing potential $\Delta(r)$are consistently solved and shown in
Fig.\ref{fig:DENSITY}. In this example, $\Delta(0)=7.2 \hbar \omega$
represents a modest strong coupling. Comparing with the
non-interacting system (dashed line), we see that the particles are
significantly dragged towards the center of the trap due to the
attractive interaction.

After solving the integral eigenvalue equations for $h_{nl}$, we
obtain the distributions of eigenvalues $|\tilde v_{nl}|^2$ and
$\tilde u^*_{nl} \tilde v_{nl}$ for various angular momentum quantum
number $l$. We note that in the case of non-interacting systems at
zero temperature, $|\tilde v_{nl}^2|$ is a step function, i.e.,
$|\tilde v_{nl}|^2= {\Theta} (\mu-\frac{3}{2}-l-2n)$. This
corresponds to the fact that non-interacting atoms fill up all the
trap energy states up to the Fermi level. For the interacting system
considered here, the sharp edge of the step function is smeared out
as shown in Fig. 2a. Such a smearing effect is similar to what
appears in uniform BCS systems \cite{deGennes}. However, the
difference here is that $f_{nl}$ are now the pairing basis instead
of plane waves.

\begin{figure}
\includegraphics [width=7 cm] {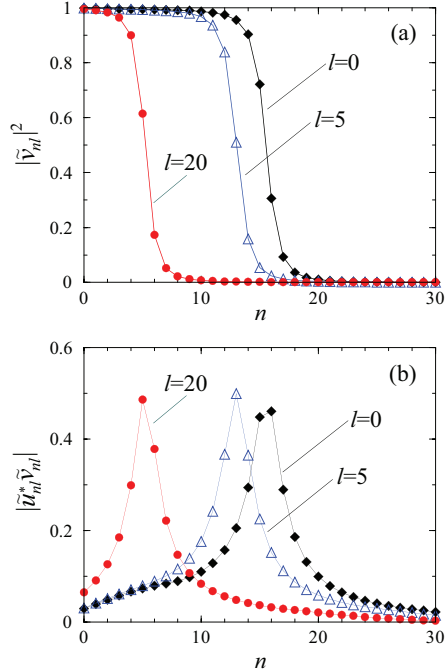}
\caption{\label{fig:OCCUPATION} (Color online) Occupation number
$|\tilde v_{kl}|^2$ (upper panel) and the pairing amplitude $\tilde
u_{nl}\tilde v_{nl}$ (lower panel) with \textit{l}=0 (diamond), 5
(triangle), 20 (circle). Same parameters as in Fig. 1.}
\end{figure}

\begin{figure}
\includegraphics [width=9.5 cm] {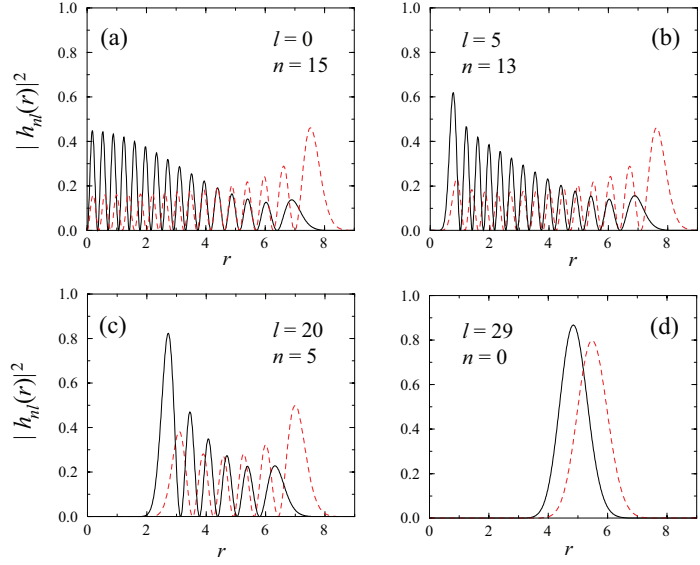}
\caption{\label{fig:MODES} (Color online) Some radial mode functions
$|h_{nl}(r)|^2$ of natural orbits, normalized to $\int |h_{nl}(r)|^2
dr =1$, with \textit{g}= -1.15 (solid line). These functions
correspond to those pairing orbits with a peak value of $\tilde
u_{nl}\tilde v_{nl}\sim 0.5$ at a given $l$. The dashed lines
correspond to the bare harmonic trap radial wave functions having
the respective quantum numbers. }
\end{figure}

In Fig. 2b, we show the quantity $\tilde u_{nl}^* \tilde v_{nl}$,
which measures the coherence between occupied and un-occupied
states. We see that $\tilde u_{nl}^* \tilde v_{nl}$ reaches a peak
at a certain values of radial quantum number $n$. These $n$'s
correspond to orbits $f_{nl}$ that have average energies close to
the chemical potential, and hence for higher $l$, the peak appears
at lower $n$. The orbits near the peak contribute most significantly
to the gap function. The shapes of some of these orbits are plotted
in Fig. \ref{fig:MODES} where the radial part $|h_{nl}(r)|^2$ at
various quantum numbers $l$ is shown. Comparing with bare
eigenfunctions of the trap (dashed line) with the same quantum
numbers $n$ and $l$, we observe similar oscillatory patterns but the
envelopes are more concentrated towards the trap center. This
indicates that the trap's eigenfunctions do not provide a good
approximation to the actual pairing orbits, at least for the
moderate strong coupling considered here. However, for a much weaker
coupling $\Delta(0) \ll \hbar \omega$ (not shown), we do find a good
agreement between pairing orbits and the bare trap's eigenfunctions
near the Fermi surface, which is expected according to the argument
in Ref. \cite{bruun}.

\begin{figure}
\includegraphics [width=8 cm] {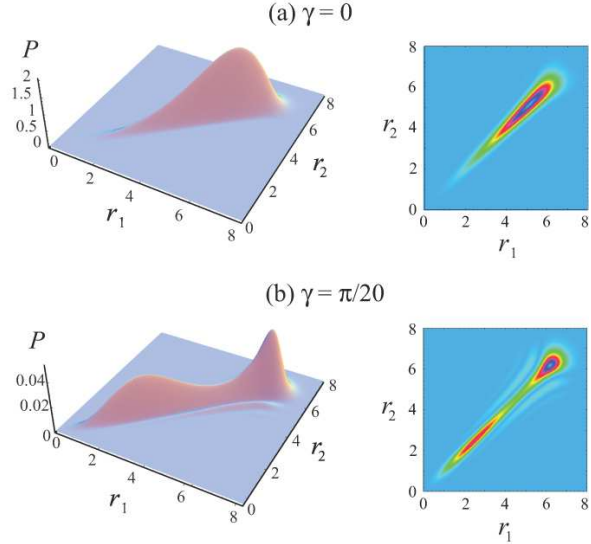}
\caption{\label{fig:COOPER} (Color online) An illustration of the
spatial profile of a Cooper pair, where
\textit{P}=$|r_1r_2F(r_1,r_2,\gamma)|^2$, with (a) $\gamma=0$ and
(b) $\gamma=\pi/20$.  Same parameters as in Fig. 1.}
\end{figure}

In Fig. \ref{fig:COOPER}, we illustrate the shape of the Cooper pair
wave function by plotting the quantity  $P=|r_1r_2
F(r_1,r_2,\gamma)|^2 $ at various angular separation $\gamma$. Note
that \textit{P} is proportional to probability density of
simultaneously finding the two particles at $\textbf{r}_1$ and
$\textbf{r}_2$, and the weighting factor $r_1r_2$ is included for
spherical systems. In both Fig. 4a and Fig. 4b, we see some
interesting fringes patterns, but the main feature is that the
function $P$ appears to be localized near the diagonal $r_1=r_2$,
indicating that both atoms are likely to be found in the same radial
distance from the trap center. In addition, by comparing Fig. 4a and
4b, $P$ drops significantly when angular separation $\gamma$
increases. This suggests that $F(r_1,r_2,\gamma)$ mainly
concentrates at $\textbf{r}_1=\textbf{r}_2$, not just the same
radial distance. For the example shown in Fig. 4a, the width of $P$
near the peak is about 1.2 which is comparable to the bare trap
ground state.

The effective Schmidt number $K$ defined in Eq. (17) is found to be
$K \approx 1749 $, which is roughly the same order of particle
number in the trap of this example. It is useful to compare this
number with the value of $K$ of a two-atom system with the same
scattering length. According to the calculation in Ref. \cite{WJ}, a
two-atom system in the ground state has a very weak entanglement if
the scattering length is small compared with the trap length unit.
For the parameters used in Fig. 4, the scattering length is
$a=-0.09$ which gives $K \approx 1$ in the two-atom system.
Therefore a Cooper pair, which is due to many-body effects,
processes a much stronger entanglement.

As a final remark, we note in Ref. \cite{law} that the ratio $K/N_j$
(where $N_\alpha=N_\beta$ is particle number of a component) is suggested to be an
indicator of determining whether a composite two-particle system
behave as a boson or not. Specifically, $K/N \gg 1$ is the regime
where composite particles can be described by creation and
annihilation operators obeying bosonic commutation relations. Here
in our example, we have $K/N \approx 1749/5456$ which is still
smaller than one. Therefore the entanglement is not strong enough to
hold the Cooper pair together as a boson.

\section{Conclusions}

To summarize, eigen-mode expansion of correlation functions provides
a powerful tool to reveal the underlying coherent structures. Such a
strategy is also known useful in various areas of physics, such as
nuclear physics and optics \cite{wolf}. In this paper, we have
examined the eigenfunctions of two-point correlation functions (3)
and (4) that can be obtained from the solutions of BdG equations.
These eigenfunctions correspond to natural pairing orbits in the BCS
state (10), and they serve as Schmidt basis vectors for the
construction of Cooper pair wave functions. Hence, one can further
analyze the quantum entanglement associated with the spatial degree
of freedom. We have demonstrated the method in a spherically trapped
system, in which the natural orbits are calculated explicitly. In
particular, our numerical results indicate features that reflect the
strong quantum entanglement between two constituent atoms in a pair.
It is useful to point out that the method could also be extended to
BEC-BCS crossover problems, where BdG equations are modified by the
molecular field and a hybrid form of BCS-molecule state may be
constructed. The structure of natural orbits in this regime is a
very interesting topic for future investigations.

\section*{Acknowledgement}
This work is supported in part by the Research Grants Council of the
Hong Kong Special Administrative Region, China (Project No. 401305
and 400504).

\appendix

\section*{Appendix}
The vanishing commutator given in Eq. (\ref{eq:COMMUTE}) can be derived under a
rather general consideration (see for example, in Ref.
\cite{dobaczewski}). Here we re-derive the result in a more
transparent way. We start with the orthogonality and completeness
relations of the quasi-particle wave functions \cite{deGennes}
\begin{eqnarray}
\int d^3r \left[u_m^*(\textbf{r})u_n(\textbf{r})+
v_m^*(\textbf{r})v_n(\textbf{r})\right]=\delta_{mn} \label{eq:ORTHO1}\\
\sum\nolimits_\eta
v_\eta(\textbf{r}_1)v_\eta^*(\textbf{r}_2)+
u_\eta^*(\textbf{r}_1)u_\eta(\textbf{r}_2)
=\delta^3(\textbf{r}_1-\textbf{r}_2). \label{eq:ORTHO2}
\end{eqnarray}
Together with Eqs. (\ref{eq:RHO_BDG}) and (\ref{eq:NU_BDG}), we can
show
\begin{eqnarray}
&&\int d^3r_1
\;\rho(\textbf{r}_1,\textbf{r})\nu(\textbf{r}_1,\textbf{r}_2) \nonumber \\
&=& \nu(\textbf{r},\textbf{r}_2)+\sum_{mn}\int d^3r_1\;u_m^*
(\textbf{r}_1)u_m(\textbf{r})u_n(\textbf{r}_1)
v_n^*(\textbf{r}_2) \nonumber \\
&=&\nu(\textbf{r},\textbf{r}_2)+\sum_{mn}u_m(\textbf{r})
v_n^*(\textbf{r}_2)\left(\delta_{mn}-\int d^3r_1\;v_m^*
(\textbf{r}_1)v_n(\textbf{r}_1) \right) \nonumber \\
&=&-\int d^3r_1\sum_{mn} u_m(\textbf{r})v_m^*(\textbf{r}_1)
v_n^*(\textbf{r}_2)v_n(\textbf{r}_1) \nonumber \\
&=& \int
d^3r_1\;\nu(\textbf{r},\textbf{r}_1)\rho(\textbf{r}_1,\textbf{r}_2).
\end{eqnarray}
Now we see that Eq. (\ref{eq:COMMUTE}) holds if the symmetric condition:
$\rho(\textbf{r}_1,\textbf{r}_2)=\rho(\textbf{r}_2,\textbf{r}_1)$ is
satisfied. With Eqs. (\ref{eq:ORTHO1}) and (\ref{eq:ORTHO2}) and the
symmetric $\rho(\textbf{r}_1,\textbf{r}_2)$, we also have the
relation:
\begin{eqnarray} \label{eq:SQUARE}
\int\rho(\textbf{r},\textbf{r}_1)\rho(\textbf{r}_1,
\textbf{r}_2)+\nu^*(\textbf{r},
\textbf{r}_1)\nu(\textbf{r}_1,\textbf{r}_2)\;d^3r_1
=\rho(\textbf{r},\textbf{r}_2).
\end{eqnarray}
Projecting both sides of Eq. (\ref{eq:SQUARE}) onto
$f_n(\textbf{r}_2)$, we obtain a relation of the eigenvalues,
\begin{eqnarray}
\lambda_n+\chi_n^2/\lambda_n=1,
\end{eqnarray}
which is consistent with the normalization condition $|\tilde
u_n|^2+|\tilde v_n|^2=1$ of BCS wave function.

\end{document}